\documentclass[letterpaper,twocolumn,aps,prl,showpacs,floatfix]{revtex4}
\pdfoutput=1

\usepackage{color,graphicx}
\usepackage{verbatim}

\usepackage[latin1]{inputenc}
\usepackage{color}
\usepackage{amsmath,amssymb,amsfonts}
\usepackage{subfigure}

\begin{document}
\DeclareGraphicsExtensions{.ps,.pdf,.eps}
\title{The Kinetic Activation-Relaxation Technique: A Powerful
 Off-lattice On-the-fly Kinetic Monte Carlo Algorithm}

\author{\firstname{Fedwa} \surname{El-Mellouhi}}
   \email{f.el.mellouhi@umontreal.ca}
   \affiliation{D\'{e}partement de Physique and Regroupement Qu\'{e}b\'{e}cois
   sur les Mat\'{e}riaux de Pointe (RQMP), Universit\'{e} de Montr\'{e}al, C.P.
   6128, Succursale Centre-Ville,\\Montr\'{e}al, Qu\'{e}bec, Canada H3C 3J7}

\author{\firstname{Normand} \surname{Mousseau}}
   \email{normand.mousseau@umontreal.ca}
   \affiliation{D\'{e}partement de Physique and Regroupement Qu\'{e}b\'{e}cois
   sur les Mat\'{e}riaux de Pointe (RQMP), Universit\'{e} de Montr\'{e}al, C.P.
   6128, Succursale Centre-Ville,\\Montr\'{e}al, Qu\'{e}bec, Canada H3C 3J7}

\author{\firstname{Laurent J.} \surname{Lewis}}
   \email{laurent.lewis@umontreal.ca}
  \affiliation{D\'{e}partement de Physique and Regroupement Qu\'{e}b\'{e}cois
   sur les Mat\'{e}riaux de Pointe (RQMP), Universit\'{e} de Montr\'{e}al, C.P.
   6128, Succursale Centre-Ville,\\Montr\'{e}al, Qu\'{e}bec, Canada H3C 3J7}

\date{\today}

\begin{abstract}
Many materials science phenomena, such as growth and self-organisation, are
dominated by activated diffusion processes and occur on timescales that are
well beyond the reach of standard-molecular dynamics simulations. Kinetic
Monte Carlo (KMC) schemes make it possible to overcome this limitation and
achieve experimental timescales. However, most KMC approaches proceed by
discretizing the problem in space in order to identify, from the outset, a
fixed set of barriers that are used throughout the simulations, limiting the
range of problems that can be addressed. Here, we propose a more flexible
approach --- the kinetic activation-relaxation technique (k-ART) --- which
lifts these constraints. Our method is based on an off-lattice,
self-learning, on-the-fly identification and evaluation of activation
barriers using ART and a topological description of events. The validity
and power of the method are demonstrated through the study of vacancy
diffusion in crystalline silicon.
\end{abstract}

\maketitle

Many problems in condensed matter and materials science involve stochastic
processes associated with the diffusion of atoms over barriers that are high
with respect to temperature and therefore inherently slow under ``normal''
conditions. Because the associated rates are small, these processes may be
considered independent; neglecting the thermal motion of atoms, it is thus
possible to deal with them using the kinetic Monte Carlo (KMC) algorithm, a
stochastic approach proposed by Bortz \emph{et al.}~\cite{Bor75, fichthorn91,
Vot02,Vot06} and based on transition state theory, whereby the evolution of a
system is determined by a set of pre-specified diffusion mechanisms, i.e.,
whose energy barriers are known beforehand. In KMC simulations, the timescale is
determined by the fastest activated processes and, in practice, timescales of
ms or longer can be reached --- much longer than accessible in traditional
molecular-dynamics (MD) simulations.

While KMC has been extensively and successfully used over the past 20 years, it
suffers from a number of drawbacks. In particular, the systems investigated must be
discretized and mapped onto a fixed lattice in order to define the various diffusion
mechanisms that need to be considered at a given moment~\cite{Vot02}. Once all
processes on the lattice have been identified (and their barriers evaluated) {\em a
priori}, the simulations simply consist in operating a diffusion event picked at
random, updating the list of possible moves in the new configuration, and iterating
this procedure long enough to cover the relevant physical timescales. This approach
works very well for simple problems (e.g., surface diffusion, metal-on-metal growth)
but fails when the systems undergo significant lattice deformations or when
long-range elastic effects are important. There have been numerous efforts to lift
these limitations, most solutions falling into one of two categories: introduction of
continuum approximations for the long-range strain deformations, and on-the-fly
evaluation of the energy barriers. The first category retains the lattice formulation
but adds long-range contributions --- which can be computed through various
extrapolation schemes --- to the barriers~\cite{mason04,sinno07}. With the second
class of solutions, there is no need to set-up a catalog of all possible activation
mechanisms. In a recently proposed self-learning KMC approach, Trushin {\em et al.}
introduced an on-the-fly search for barriers but displacements were restricted to be
on-lattice\cite{trushin05}. In other cases, a limited number of activated events
using the the ART-like dimer~\cite{Hen99,hontinfinde06} or eigenvector-following
methods~\cite{middleton04} are generated at each step in order to construct a small
catalog which serves to determine the next move. Thus, these
methods~\cite{mason04,sinno07,trushin05} are still limited by the lattice description
of the problem and the approximate character of the elastic energies.
On-the-fly/off-lattice approaches ~\cite{Hen99,hontinfinde06,middleton04}, on the
other hand, while more flexible, are currently inefficient as they do not take
advantage of the knowledge of previously-encountered events, and are therefore only
useful for small systems with very few barriers.

In this Letter, we introduce a powerful on-the-fly/off-lattice KMC method which
achieves speed-ups as large as 4000 over standard MD for complex systems, while
retaining a complete description of the relevant physics, including long-range
elastic interactions. Our approach is based on the activation-relaxation technique
(ART nouveau)~\cite{Bar96, Malek00} for generating events and calculating barriers;
the gain in efficiency is achieved through a topological classification of atomic
environments, which allows configurations and events to be recognized and stored
efficiently, and used again as the simulation proceeds, i.e., the method is
self-learning. We demonstrate the validity and efficiency of this kinetic ART (k-ART)
approach by applying it to the problem of vacancy diffusion in crystalline silicon
and comparing to full MD simulations.

\begin{figure}
        \centering
                \includegraphics[ height=5.5cm, angle=0]{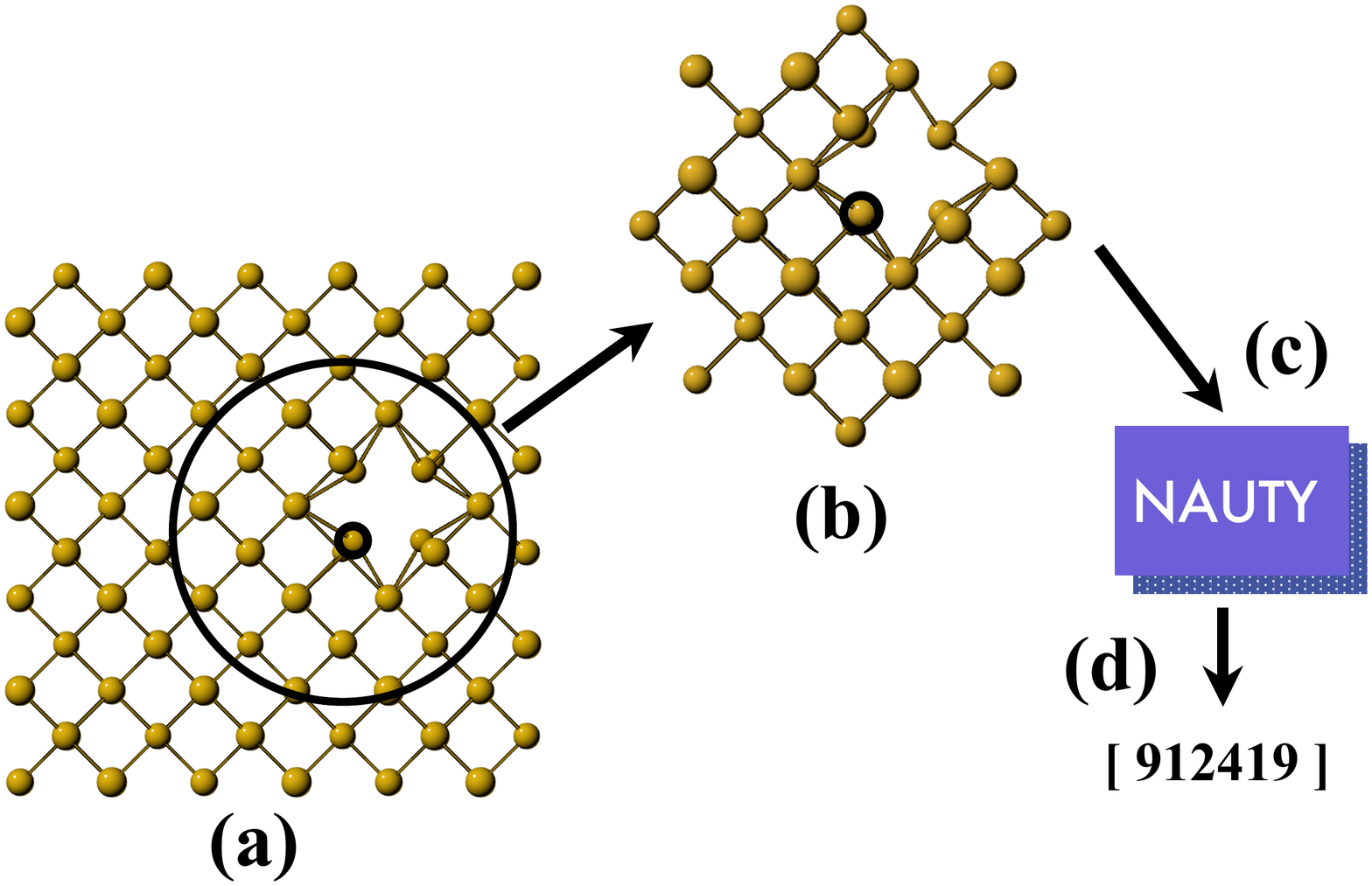}
        \caption{(Color online) Local topology analysis in k-ART: a truncated graph (b) is
        extracted from the complete lattice around the highlighted central atom (a), and analysed using {\tt
        nauty} (c) which returns a unique key associated with the given topology
        (d).}
        \label{fig:topo}
\end{figure}

Before describing k-ART, it is useful to discuss the topological characterization.
For each configuration, a connectivity graph formed by the network of local neighbors
is first constructed. These may correspond to covalently bonded atoms in
semiconductors, or faces in the Vorono\"{\i} tesselation of compact materials. It is
important that the configuration be uniquely defined through this network, i.e., the
connectivity graph must lead to a unique structure once relaxed with a given
interatomic potential. In order to classify the activated processes, a truncated
graph is constructed around each atom, as illustrated Fig.~\ref{fig:topo}, the
size of which depends on the physics of the system under study. In the case of Si,
for example, we define the local environnement around an atom by a sphere of radius
5.0 \AA, which includes about 40 atoms; two neighbours are bonded if their distance
is less than 2.8 \AA. An event is defined as a change in the topology of the local
graph. This classification is performed using the freely available topological
software {\tt nauty}, developed by McKay\cite{nauty}, which provides the topology
index and all information necessary for uniquely identifying each environment,
including the permutation key needed to reconstruct a specific geometry from the
generic topology and a set of reference positions.

Events which have been learned are stored for subsequent use; in practice, the atomic
positions of the initial state as well as the associated topologies for the initial,
transition and final truncated graphs are saved in memory. If needed, the transition
and final state configurations may be reconstructed from the reference geometry
through a series of symmetry operations extracted from the topological analysis. This
results in a considerable reduction in the amount of data that needs to be generated
and manipulated. For a single vacancy in c-Si, for example, only 20 different
topologies are necessary to describe all possible local environments, irrespective of
the system size. Moreover, as the system evolves and previously-encountered
topologies are recognized, it is only necessary to update the table of active events,
the cost of which is negligible as we will see below.

We now turn to a detailed description of the k-ART algorithm. Starting from an
initial relaxed configuration, the various local topologies are characterized with
{\tt nauty} and, for each topology, possible events are constructed with ART
nouveau~\cite{Bar96, Malek00}, which has been shown to efficiently identify the
relevant diffusion mechanism in a wide range of systems, either crystalline or
amorphous, with both empirical and \emph{ab initio} methods~\cite{Song_NEBM, Valiq03,
Elm04, Mal07}. Within this approach, the configuration is slowly pushed along a
randomly selected direction until an unstable direction appears in the Hessian; this
is followed while minimising the energy in the perpendicular hyperplane until the
system converges onto a saddle point and the system is then pushed over the barrier
and relaxed into a new minimum. Since activated processes involve only a finite
number of atoms, each event is initiated by displacing a given atom and its
neighbours within a small, local region in a random direction. The exact size of the
displacement regions depends on the system under study; in semiconductors, they
typically involve first and second nearest-neighbours. The initial convergence
criterion for the saddle point search is set to 1.0 eV/\AA\ in order to accelerate
convergence (but see below).
\begin{figure}
        \centering
                \includegraphics[height=8cm, angle=0]{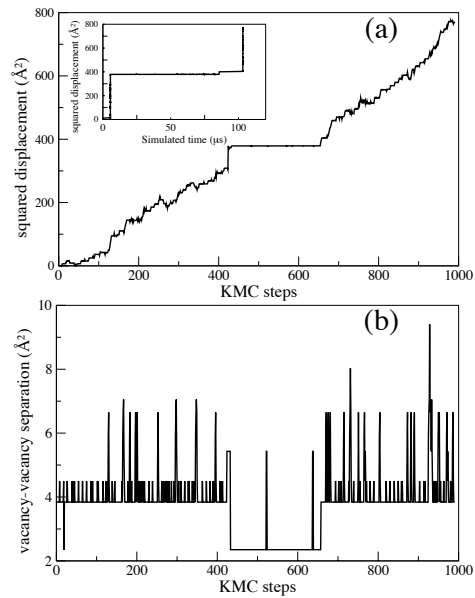}
        \caption{(a) Total squared displacement as function of KMC steps (or
        time, in the inset). (b) Distance between the two vacancies as a
        function of KMC steps. }
        \label{fig:bila}
\end{figure}

To simplify labelling, each event is assigned to the topology centered on the
atom that moved the most during the event, irrespective of the initial trial
assignment. The events are stored as displacement vectors from the reference
state to the transition and the final states; these are used to reconstruct
all specific events associated with a given topology throughout the lattice.
Once the list of topologies and associated barriers is set (or has been
updated), all low-energy events (which we define for Si as having barriers of
15 $k_BT$ or less) are reconstructed from the topology and re-relaxed with a
stricter convergence criterion of 0.1 eV/\AA\ in order to accurately take
into account the local environment and the long-range interactions, leading
to a precision of about 0.01 eV on the barrier height. At this
point, two types of events are in the catalog: (1) ``generic'' events, that
include all high-energy barriers, and (2) ``specific'' events, where all
low-energy barriers, dominating the kinetics, are relaxed individually. We
associate a transition rate $r_i = \tau_0 \exp\left(\Delta E_i/k_BT\right)$
to each event, where $\tau_0$ is fixed at the outset and, for simplificity,
assumed to be the same ($=10^{13} s^{-1}$) for all events.
From this list, and following Bortz \emph{et al.}~\cite{Bor75}, the elapsed
time to the next event is computed as $\Delta t = -\ln \mu / \sum_i r_i$,
with $\mu$ a random number in the $[0,1[$ interval. Finally, an event is
selected with a weight proportional to its rate and is operated; the clock is
pushed forward and the process starts again: The topology of all atoms
belonging to the local environment around the new state is constructed; if a
new topology is found, a series of ART nouveau searches are launched;
otherwise, we proceed to the next step. After all events are updated, the
low-lying barriers are, once again, relaxed before calculating the time
increment and selecting the next move.

As the system evolves, it may get trapped in a set of local configurations separated
by very low energy barriers that dominate the dynamics without yielding diffusion. An
exact solution to dealing with such ``flickers'' has been proposed by Athenes {\it et
al.}~\cite{Athenes97}, but we elect here to use a simpler limited-memory Tabu-like
approach~\cite{tabu} which proceeds by banning transitions rather than
states~\cite{chubynsky06}. In brief, at any given moment, we keep in memory (the
``memory kernel'') the $n$ previous transitions. If a planned transition is already
in memory, it is blocked and the initial or the final configuration of this move is
adopted with the appropriate Boltzmann probability; the transition is also blocked
for the next $n$ jumps and removed from the list of possible events. As was shown in
Ref.~\cite{chubynsky06}, this approach is thermodynamically exact and is kinetically
valid as long as the memory is short compared to the timeline of evolution of the
system.
\begin{figure}
        \centering
                \includegraphics[height=6cm, angle=0]{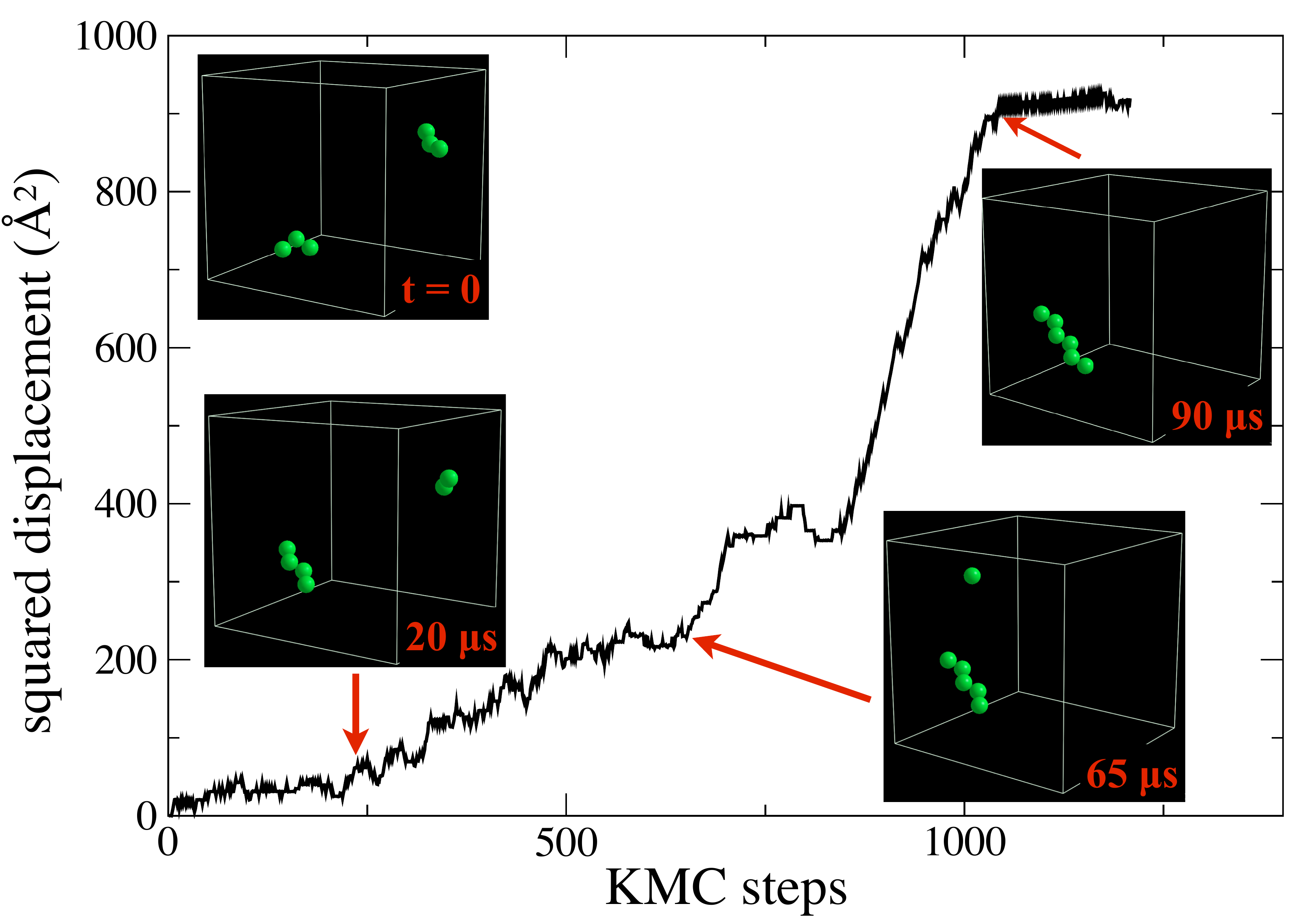}
        \caption{(Color online) Total squared displacement as function of KMC steps for the
        6-vacancy system.}
        \label{fig:diff}
\end{figure}

We now demonstrate the validity and efficiency of our method by studying the
diffusion of systems of two and six vacancies in a 1000-atom Stillinger-Weber
c-Si sample. For the 2-vacancy system, we start by removing two
second-neighbour Si atoms, then perform a k-ART run for 200 CPU hours on a
single 1.5 GHz Itanium 2 processor. During this time, the vacancies perform
about 1000 jumps, corresponding to a diffusion time of about 100 $\mu$s.
Figure~\ref{fig:bila}~(a) shows the total squared displacement of the atoms
as a function of KMC steps (i.e., events) and, in the inset, effective time.
Two types of behaviour are clearly visible. During the first 400 steps ($\sim
4$ $\mu$s), diffusion takes place through correlated single vacancy hops, the
two vacancies maintaining a separation oscillating between 3.85 and 4.5 \AA. At
about step 400, the two vacancies become trapped as a single divacancy,
characterized by small local rearrangements, and remain so for about 80
$\mu$s before partially breaking apart and resuming its 2-vacancy correlated
walk. The correlated motion is best seen in Fig.~\ref{fig:bila}~(b), where we
plot the distance between the two vacancies as a function of KMC steps : the
two vacancies remain bound in a first or second-neighbour state for the whole
simulation, except for occasional excursions to larger distances.
This striking result illustrates perfectly the strength of k-ART in handling
fully the impact of lattice deformation on diffusion, which here induces the
two vacancies to return rapidly to a tightly-correlated configuration.

For the 6-vacancy problem, now, we start with a configuration containing two
3-vacancy clusters placed far away from each other, as shown in the $t=0$ snapshot in
Fig.~\ref{fig:diff}. This configuration is challenging because the dynamics is
dominated by a series of local rearrangements and reorientations associated with
low-energy barriers that preserve the compactness of the cluster, i.e., breaking it
appart is very difficult. To test this, we first ran a 30 ns MD simulation at 500 K;
no dissociation or diffusion events were observed. Likewise, nothing happened in a
5000-step k-ART simulation {\em without} memory kernel, which covered 8 $\mu$s. These
two calculations required roughly the same computational effort; we thus already
conclude that k-ART is at least 250 times faster than MD; this is fast, but we can do
much better by invoking the memory kernel to eliminate the flicker problem which is
inherent to such complex materials.

Thus, we carried out a third simulation of the 6-vacancy problem using k-ART
{\em and} the memory kernel. Figure~\ref{fig:diff} shows the squared
displacement as function of KMC steps. We observe, in agreement with the
previous two simulations, that the initial state is fairly stable: the system
flickers during the first 20 $\mu$s (160 KMC steps), in agreement with the MD and the k-ART simulation without memory kernel. At 20 $\mu$s,
one vacancy breaks away from the top right cluster and quickly moves to the
other cluster, forming a 4-vacancy chain and leaving a divacancy behind. As
in the 2-vacancy simulation, this divacancy diffuses through the box for
about 45 $\mu$s (525 KMC steps) as a correlated pair. Finally, at event 685 (65
$\mu$s), the divacancy breaks appart and one vacancy rapidly joins the larger
cluster. The lone monovacancy diffuses through the lattice during 25 $\mu$s
and eventually joins the 5-vacancy cluster, forming a stable hexavacancy
chain with a total energy 2 eV lower than that of the initial configuration.
Not surprisingly, for the last 150 KMC steps (20 $\mu$s), the diffusion
becomes negligible and we observe only unsuccessful attempts to dissociate
the hexavacancy cluster.

\begin{figure}
        \centering
                \includegraphics[height=6cm, angle=0]{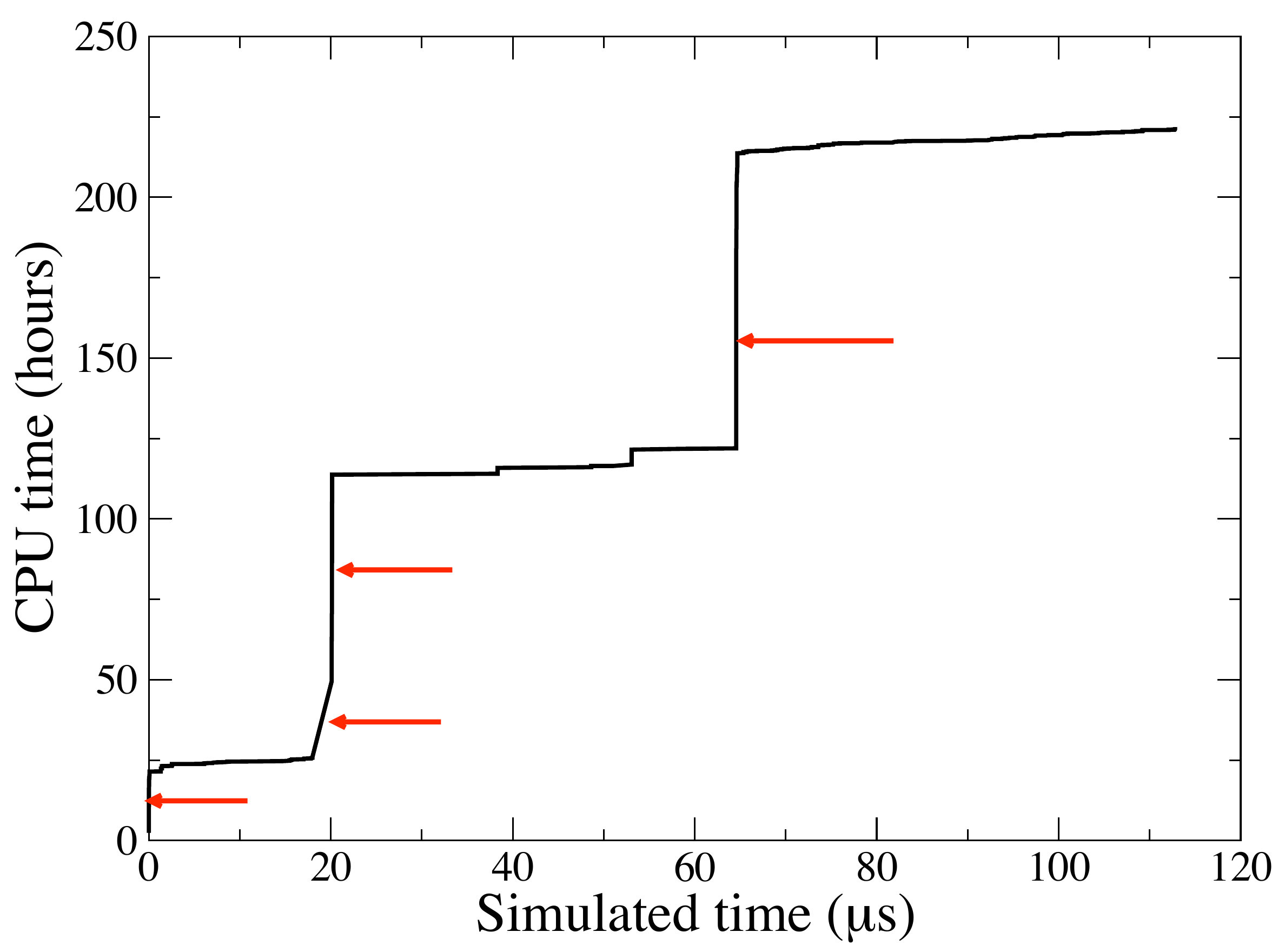}
        \caption{(Color online) CPU time versus simulated time for the
        hexavacancy aggregation problems. Red arrows indicate extensive
        self-learning phases in k-ART}
        \label{fig:CPU}
\end{figure}

In terms of efficiency, k-ART with the memory kernel is about 4000 times
faster than MD, with 110 $\mu$s simulated in 220 CPU hours. In k-ART most of
the computational time is spent in identifying events associated with new
topologies. This is clear in Figure~\ref{fig:CPU} where we plot CPU time
versus simulated time for k-ART, with the learning phases indicated by
arrows. Sampling is considerable: at the end of this run, 17237 different
events were generated, associated with 1964 initial topologies, for an
average of almost 9 events per topology; at each step during the simulation,
the system presents about 80 to 120 different topologies. Since each atom is
associated with a topology, about 9000 different barriers are considered at
each KMC step.

While the cost of k-ART is significantly higher than lattice KMC, it is
ideally suited to problems that have been left aside until now because of the
importance of off-lattice positions, surface reconstruction and long-range
elastic effects, as we have shown in the study of 2 and 6 Si vacancy
diffusion. Because the method is inherently local, a number of improvements
can be envisaged that will yield considerable acceleration to the code.
Parallelizing the management of events and barriers, for example, should
speed up the calculations by a factor of 10 or 20. Moreover, because of the
topological classification, the catalog of events may be stored and reused at
a later time, thus accelerating new simulations.
Kinetic ART is an exciting self-learning, off-lattice kinetic Monte-Carlo
algorithm that opens the door to the numerical study of problems such as
semiconductor growth, self-organization, defect diffusion and interface
mixing, that have until now been out of the reach of simulations.


\begin{acknowledgments}

This work has been supported by the Canada Research Chairs program and by
grants from the Natural Sciences and Engineering Research Council of Canada
(NSERC) and the \textit{Fonds Qu\'eb\'ecois de la Recherche sur la Nature et
les Technologies} (FQRNT). We are grateful to the \textit{R\'eseau
Qu\'eb\'ecois de Calcul de Haute Performance} (RQCHP) for generous allocations
of computer resources.

\end{acknowledgments}


\begin{thebibliography}{20}
\expandafter\ifx\csname natexlab\endcsname\relax\def\natexlab#1{#1}\fi
\expandafter\ifx\csname bibnamefont\endcsname\relax
  \def\bibnamefont#1{#1}\fi
\expandafter\ifx\csname bibfnamefont\endcsname\relax
  \def\bibfnamefont#1{#1}\fi
\expandafter\ifx\csname citenamefont\endcsname\relax
  \def\citenamefont#1{#1}\fi
\expandafter\ifx\csname url\endcsname\relax
  \def\url#1{\texttt{#1}}\fi
\expandafter\ifx\csname urlprefix\endcsname\relax\def\urlprefix{URL }\fi
\providecommand{\bibinfo}[2]{#2}
\providecommand{\eprint}[2][]{\url{#2}}

\bibitem[{\citenamefont{Bortz and Kalos}(1975)}]{Bor75}
\bibinfo{author}{\bibfnamefont{A.~B.} \bibnamefont{Bortz}} \bibnamefont{and}
  \bibinfo{author}{\bibfnamefont{M.~H.} \bibnamefont{Kalos}},
  \bibinfo{journal}{J. Comput. Phys.} \textbf{\bibinfo{volume}{178}},
  \bibinfo{pages}{10} (\bibinfo{year}{1975}).

\bibitem[{\citenamefont{Fichthorn and Weinberg}(1991)}]{fichthorn91}
\bibinfo{author}{\bibfnamefont{K.}~\bibnamefont{Fichthorn}} \bibnamefont{and}
  \bibinfo{author}{\bibfnamefont{W.}~\bibnamefont{Weinberg}},
  \bibinfo{journal}{J Chem Phys} \textbf{\bibinfo{volume}{95}},
  \bibinfo{pages}{1090} (\bibinfo{year}{1991}).

\bibitem[{\citenamefont{Voter et~al.}(2002)\citenamefont{Voter, Montalenti, and
  Germann}}]{Vot02}
\bibinfo{author}{\bibfnamefont{A.~F.} \bibnamefont{Voter}},
  \bibinfo{author}{\bibfnamefont{F.}~\bibnamefont{Montalenti}},
  \bibnamefont{and} \bibinfo{author}{\bibfnamefont{T.~C.}
  \bibnamefont{Germann}}, \bibinfo{journal}{Annu. Rev. Mater. Res.}
  \textbf{\bibinfo{volume}{34}}, \bibinfo{pages}{321} (\bibinfo{year}{2002}).

\bibitem[{\citenamefont{Voter}(2007)}]{Vot06}
\bibinfo{author}{\bibfnamefont{A.~F.} \bibnamefont{Voter}},
  \emph{\bibinfo{title}{Radiation Effects in Solids. ed.\ by K.E. Sickafus,
  E.A. Kotomin, and B.P. Uberuaga}} (\bibinfo{publisher}{Springer, NATO
  Publishing Unit, Dordrecht, The Netherlands}, \bibinfo{year}{2007}), vol.
  \bibinfo{volume}{235}, chap. \bibinfo{chapter}{Introduction to the Kinetic
  Monte Carlo Method}, pp. \bibinfo{pages}{1--24}.

\bibitem[{\citenamefont{Mason et~al.}(2004)\citenamefont{Mason, Rudd, and
  Sutton}}]{mason04}
\bibinfo{author}{\bibfnamefont{D.}~\bibnamefont{Mason}},
  \bibinfo{author}{\bibfnamefont{R.}~\bibnamefont{Rudd}}, \bibnamefont{and}
  \bibinfo{author}{\bibfnamefont{A.}~\bibnamefont{Sutton}}, \bibinfo{journal}{J
  Phys-Condens Mat} \textbf{\bibinfo{volume}{16}}, \bibinfo{pages}{S2679}
  (\bibinfo{year}{2004}).

\bibitem[{\citenamefont{Sinno}(2007)}]{sinno07}
\bibinfo{author}{\bibfnamefont{T.}~\bibnamefont{Sinno}}, \bibinfo{journal}{J
  Cryst Growth} \textbf{\bibinfo{volume}{303}}, \bibinfo{pages}{5}
  (\bibinfo{year}{2007}).

\bibitem[{\citenamefont{Trushin et~al.}(2005)\citenamefont{Trushin, Karim,
  Kara, and Rahman}}]{trushin05}
\bibinfo{author}{\bibfnamefont{O.}~\bibnamefont{Trushin}},
  \bibinfo{author}{\bibfnamefont{A.}~\bibnamefont{Karim}},
  \bibinfo{author}{\bibfnamefont{A.}~\bibnamefont{Kara}}, \bibnamefont{and}
  \bibinfo{author}{\bibfnamefont{T.}\bibfnamefont{S.}~\bibnamefont{Rahman}},
  \bibinfo{journal}{Phys Rev B} \textbf{\bibinfo{volume}{72}},
  \bibinfo{pages}{115401} (\bibinfo{year}{2005}).

\bibitem[{\citenamefont{Henkelman and J\'onsson}(1999)}]{Hen99}
\bibinfo{author}{\bibfnamefont{G.}~\bibnamefont{Henkelman}} \bibnamefont{and}
  \bibinfo{author}{\bibfnamefont{J.}~\bibnamefont{J\'onsson}},
  \bibinfo{journal}{J.\ Chem.\ Phys.} \textbf{\bibinfo{volume}{111}},
  \bibinfo{pages}{7010} (\bibinfo{year}{1999}).

\bibitem[{\citenamefont{Hontinfinde et~al.}(2006)\citenamefont{Hontinfinde,
  Rapallo, and Ferrando}}]{hontinfinde06}
\bibinfo{author}{\bibfnamefont{F.}~\bibnamefont{Hontinfinde}},
  \bibinfo{author}{\bibfnamefont{A.}~\bibnamefont{Rapallo}}, \bibnamefont{and}
  \bibinfo{author}{\bibfnamefont{R.}~\bibnamefont{Ferrando}},
  \bibinfo{journal}{Surf Sci} \textbf{\bibinfo{volume}{600}},
  \bibinfo{pages}{995} (\bibinfo{year}{2006}).

\bibitem[{\citenamefont{Middleton and Wales}(2004)}]{middleton04}
\bibinfo{author}{\bibfnamefont{T.}~\bibnamefont{Middleton}} \bibnamefont{and}
  \bibinfo{author}{\bibfnamefont{D.}~\bibnamefont{Wales}}, \bibinfo{journal}{J
  Chem Phys} \textbf{\bibinfo{volume}{120}}, \bibinfo{pages}{8134}
  (\bibinfo{year}{2004}).

\bibitem[{\citenamefont{Barkema and Mousseau}(1996)}]{Bar96}
\bibinfo{author}{\bibfnamefont{G.~T.} \bibnamefont{Barkema}} \bibnamefont{and}
  \bibinfo{author}{\bibfnamefont{N.}~\bibnamefont{Mousseau}},
  \bibinfo{journal}{Phys.\ Rev.\ Lett.} \textbf{\bibinfo{volume}{77}},
  \bibinfo{pages}{4358} (\bibinfo{year}{1996}).

\bibitem[{\citenamefont{Malek and Mousseau}(2000)}]{Malek00}
\bibinfo{author}{\bibfnamefont{R.}~\bibnamefont{Malek}} \bibnamefont{and}
  \bibinfo{author}{\bibfnamefont{N.}~\bibnamefont{Mousseau}},
  \bibinfo{journal}{Phys. Rev. E} \textbf{\bibinfo{volume}{62}},
  \bibinfo{pages}{7723} (\bibinfo{year}{2000}).

\bibitem[{\citenamefont{McKay}(1981)}]{nauty}
\bibinfo{author}{\bibfnamefont{B.~D.} \bibnamefont{McKay}},
  \bibinfo{journal}{Congressus Numerantium} \textbf{\bibinfo{volume}{30}},
  \bibinfo{pages}{45} (\bibinfo{year}{1981}).

\bibitem[{\citenamefont{Song et~al.}(2000)\citenamefont{Song, Malek, and
  Mousseau}}]{Song_NEBM}
\bibinfo{author}{\bibfnamefont{Y.}~\bibnamefont{Song}},
  \bibinfo{author}{\bibfnamefont{R.}~\bibnamefont{Malek}}, \bibnamefont{and}
  \bibinfo{author}{\bibfnamefont{N.}~\bibnamefont{Mousseau}},
  \bibinfo{journal}{Phys. Rev. B} \textbf{\bibinfo{volume}{62}},
  \bibinfo{pages}{15680} (\bibinfo{year}{2000}).

\bibitem[{\citenamefont{Valiquette and Mousseau}(2003)}]{Valiq03}
\bibinfo{author}{\bibfnamefont{F.}~\bibnamefont{Valiquette}} \bibnamefont{and}
  \bibinfo{author}{\bibfnamefont{N.}~\bibnamefont{Mousseau}},
  \bibinfo{journal}{Phys.\ Rev.\ B} \textbf{\bibinfo{volume}{68}},
  \bibinfo{pages}{125209} (\bibinfo{year}{2003}).

\bibitem[{\citenamefont{El-Mellouhi et~al.}(2004)\citenamefont{El-Mellouhi,
  Mousseau, and Ordej\'on}}]{Elm04}
\bibinfo{author}{\bibfnamefont{F.}~\bibnamefont{El-Mellouhi}},
  \bibinfo{author}{\bibfnamefont{N.}~\bibnamefont{Mousseau}}, \bibnamefont{and}
  \bibinfo{author}{\bibfnamefont{P.}~\bibnamefont{Ordej\'on}},
  \bibinfo{journal}{Phys.\ Rev.\ B} \textbf{\bibinfo{volume}{70}},
  \bibinfo{pages}{205202} (\bibinfo{year}{2004}).

\bibitem[{\citenamefont{Malouin et~al.}(2007)\citenamefont{Malouin,
  El-Mellouhi, and Mousseau}}]{Mal07}
\bibinfo{author}{\bibfnamefont{M.-A.} \bibnamefont{Malouin}},
  \bibinfo{author}{\bibfnamefont{F.}~\bibnamefont{El-Mellouhi}},
  \bibnamefont{and} \bibinfo{author}{\bibfnamefont{N.}~\bibnamefont{Mousseau}},
  \bibinfo{journal}{Phys. Rev. B} \textbf{\bibinfo{volume}{76}}
  (\bibinfo{year}{2007}).

\bibitem[{\citenamefont{Athenes et~al.}(1997)\citenamefont{Athenes, Bellon, and
  Martin}}]{Athenes97}
\bibinfo{author}{\bibfnamefont{M.}~\bibnamefont{Athenes}},
  \bibinfo{author}{\bibfnamefont{P.}~\bibnamefont{Bellon}}, \bibnamefont{and}
  \bibinfo{author}{\bibfnamefont{G.}~\bibnamefont{Martin}},
  \bibinfo{journal}{Phil Mag A} \textbf{\bibinfo{volume}{76}},
  \bibinfo{pages}{565} (\bibinfo{year}{1997}).

\bibitem[{\citenamefont{Glover and Laguna}(1997)}]{tabu}
\bibinfo{author}{\bibfnamefont{F.}~\bibnamefont{Glover}} \bibnamefont{and}
  \bibinfo{author}{\bibfnamefont{M.}~\bibnamefont{Laguna}},
  \emph{\bibinfo{title}{Tabu Search}} (\bibinfo{publisher}{Kluwer Academics},
  \bibinfo{address}{Dordrecht}, \bibinfo{year}{1997}).

\bibitem[{\citenamefont{Chubynsky et~al.}(2006)\citenamefont{Chubynsky, Vocks,
  Barkema, and Mousseau}}]{chubynsky06}
\bibinfo{author}{\bibfnamefont{M.~V.} \bibnamefont{Chubynsky}},
  \bibinfo{author}{\bibfnamefont{H.}~\bibnamefont{Vocks}},
  \bibinfo{author}{\bibfnamefont{G.~T.} \bibnamefont{Barkema}},
  \bibnamefont{and} \bibinfo{author}{\bibfnamefont{N.}~\bibnamefont{Mousseau}},
  \bibinfo{journal}{J Non-Cryst Solids} \textbf{\bibinfo{volume}{352}},
  \bibinfo{pages}{4424} (\bibinfo{year}{2006}).

\end{thebibliography}
\end{document}